Rail. Eng. Science
https://doi.org/10.1007/s40534-020-00213-y

# Real-time multibody modeling and simulation of a scaled bogie test rig

Sundar Shrestha[1,2] · Maksym Spiryagin[1,2] · Qing Wu[1,2]Received: 13 March 2020 / Revised: 9 May 2020 / Accepted: 12 May 2020
© The Author(s) 2020**Abstract** In wheel–rail adhesion studies, most of the test rigs used are simplified designs such as a single wheel or wheelset, but the results may not be accurate. Alternatively, representing the complex system by using a full vehicle model provides accurate results but may incur complexity in design. To trade off accuracy over complexity, a bogie model can be the optimum selection. Furthermore, only a real-time model can replicate its physical counterpart in the time domain. Developing such a model requires broad expertise and appropriate software and hardware. A few published works are available which deal with real-time modeling. However, the influence of the control system has not been included in those works. To address these issues, a real-time scaled bogie test rig including the control system is essential. Therefore, a 1:4 scaled bogie roller rig is developed to study the adhesion between wheel and roller contact. To compare the performances obtained from the scaled bogie test rig and to expand the test applications, a numerical simulation model of that scaled bogie test rig is developed using Gensys multibody software. This model is the complete model of the test rig which delivers more precise results. To exactly represent the physical counterpart system in the time domain, a real-time scaled bogie test rig (RT-SBTR) is developed after four consecutive stages. Then, to simulate the RT-SBTR to solve the internal state equations and functions representing the physical counterpart system in equal or less than actual time, the real-time simulation environment is prepared in two stages. To such end, the computational time improved from 4 times slower than real time to 2 times faster than real time. Finally, the real-time scaled bogie model is also incorporated with the braking control system which slightly reduces the computational performances without affecting real-time capability.

**Keywords** Bogie modeling · Scaled bogie test rig · Real-time simulation · Wheel–rail adhesion · Software in loop## 1 Introduction

Railway vehicles are the most efficient land-based transportation, generally having a higher carrying capacity at a comparatively low energy cost. To progressively achieve the requirements of power, speed, and complexity, the modern railway vehicle requires more advances in different domains of railway operation to test, validate, and optimize performance. To perform such test and validation, the test rig has become more widespread because it provides a low cost, highly repeatable, low maintenance, and very safe modular testing environment.

In the development of mechatronics product or system, for instance, a test rig integrates various stages such as design, verification of design (simulation), and physical implementation. It is important to understand the system development process to avoid the huge cost incurred in later modification. Additionally, the process focuses on the safety and reliability of the system. Figure 1 shows the stages to develop the scaled test rig in the laboratory. The

✉ Sundar Shrestha
s.shrestha@cqu.edu.au

1 Centre for Railway Engineering, Central Queensland University, Rockhampton, Australia
2 Rail Manufacturing Cooperative Research Centre (CRC), Melbourne, VIC, AustraliaPublished online: 12 June 2020Springer

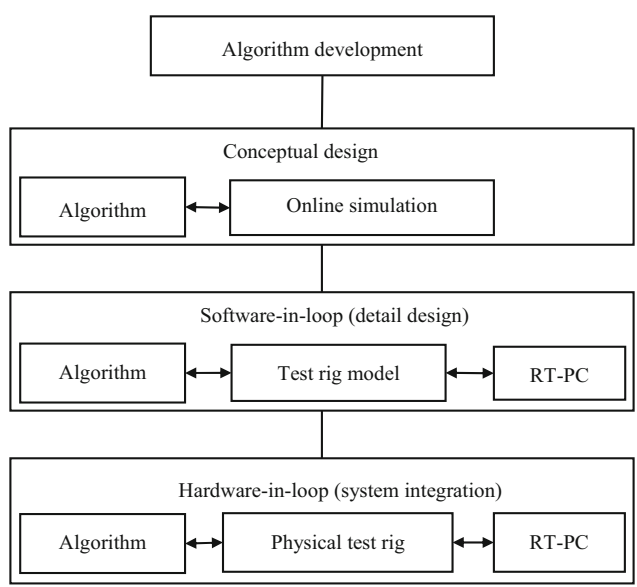

Fig. 1 Development stages of the scaled bogie test rig

algorithm development phase and conceptual design phase have been accomplished and published in the literature [1]. In this paper, the software-in-loop simulation stage of the scaled test rig development is explained. A simulation model is the replication of the physical counterpart, which is developed by using specialized multibody software (MBS). It is essential for such a model to compare the performances obtained from the real vehicle and/or test rigs that are imitating the vehicle behavior. The application of test rigs to the study of railway vehicle system dynamics has become more widespread because it provides a low cost, highly repeatable, low maintenance, and very safe modular testing environment.

### 1.1 Research topic 1

Before modeling and designing a test rig, it is necessary to answer a vital question, "What is the optimal design of the test rig?". The devices and validation technologies that are applied for adhesion studies can be classified into two types by their size, full-scaled size and reduced-scaled size. The full-scale test rigs are suitable to assess the dynamic performances of a prototype vehicle with high accuracy. Their higher cost is the main concern that limits their extensive use. On the other hand, reduced-scale test rigs offer numerous advantages, namely that they occupy less space, and they are easier to maintain, easier to handle and require less effort to change several vehicle parameters. Hence, the reduced-scale version of the test rig was selected for this study. Nevertheless, the benefits of this are offset by some adverse factors. One of the major problematic factors is scaling. This problem has been investigated by many authors as canvassed in [2, 3]. In [2], the authors have mentioned three different scaling strategies, namely DLR, INRETS, and MMU. Furthermore, scholars [4, 5] have compared three scaling methods proposed by Pascal, Iwnicki, and Jaschinski. These investigations report two fundamental strategies of scaling, namely dimension-based scaling (i.e., geometrical scaling) and the equation of motion-based scaling. However, neither of these strategies works when dealing with some parameters which are not scalable such as material elasticity and acceleration due to gravity. The scaling strategies implemented in this project are explained in [2].

Furthermore, these reduced-scale and full-scale test rigs can both be further categorized into four subclasses according to the type of the sub-system being used. They are the single wheel, wheelset, bogie, and full vehicle. The presence of a higher level of a sub-system in the test rig increases the accuracy of the test results; however, it also raises the complexity. Most of the test rigs used are single wheel and single wheelset which are simple in design, but there is room for development [6]. Conversely, the complex system such as a full vehicle model delivers better results but may incur complexity in design. To trade off accuracy over complexity, the optimum selection can be a bogie model. While most of the published papers consider single wheelset dynamics for adhesion studies [7–12], this paper will consider bogie dynamics for adhesion studies and contribute to the knowledge in that particular domain.

### 1.2 Research topic 2

The second research question is "Is it possible to develop a real-time test rig model with braking control system integrated?". Currently, simulation modeling has been extensively used to study the different domains of railway operation. The advanced simulation modeling to accurately characterize vehicle–track dynamics of the railway is performed in [13–15], while [16–23] show the simulation modeling of railway under traction and/or braking. Moreover, the exact representation of a physical system and its performance in the time domain is a challenging task. Only a real-time model can replicate its physical counterpart in the time domain. In a real-time simulation, the simulation is performed in a discrete-time solver with a fixed time step by solving the internal state equations and functions representing the physical counterpart system in less than that fixed step. It is useful in the design phase to investigate controller/actuators before implementing these into the physical model. This allows significant experimental cost reduction and examines the new/potential designs in less time. However, developing a real-time model and integrating them into multidisciplinary models involve broad expertise as well as appropriate software and hardware.

Since the real-time model is the replication of the physical system using specialized multibody system dynamics software running on a computer, one of the major requirements of such a software package is the computational speed which needs to be faster than real time. In the rail vehicle dynamics domain, no such software package has been found so far.

The possibilities of developing the model for real-time application in MATLAB/Simulink have been shown by Bosso et al. [24]. In [18], researchers developed a real-time model by transferring the model developed in SimMechanics (toolbox of Simulink) into the real-time workshop in the dSPACE platform. MATLAB models integrally possess difficulties with modifying the parameters. Additionally, the nonlinear components were characterized by the comparative linear model, resulting in less accurate simulation in the presence of disturbances. An advanced approach to developing a real-time test rig model for a heavy haul locomotive based on the Gensys MBS package is presented and verified through co-simulation in [25]. However, the influence of the control system calculation process was not included in that model which potentially decreases the real-time performance. To address that issue, a real-time model integrating the braking control system is developed in this study that can emulate the behavior of the physical counterpart faster than the actual time.

This paper is organized as follows. In Sect. 1, the practical objectives of the investigations are mentioned. In Sect. 2, the methodology to solve those objectives/problems mentioned in Introduction section is formulated. In Sect. 3, a complete model of the bogie test rig is developed and validated with experimental results. In Sect. 4, various stages to prepare a real-time simulation environment and phases to develop the real-time scaled bogie test rig from the complete model are explained. Simulations with different case studies and their results are presented in Sect. 5. The simulation results and comparisons are further discussed and analyzed in Sect. 6.

## 2 Methodology

The methodology to address the research questions mentioned in the introduction section is presented in Fig. 2. To answer the first question, a scaled bogie test rig (SBTR) model of a railway freight wagon is developed. The model is first tested to ensure that the model code used is free of errors and that the multibody model behaves as expected when basic static and dynamic analyses are performed. The debugged model is then validated with the experimental results of its physical counterpart.

To address the second research gap, a real-time environment is essential to simulate the real-time model. The real-time environment is created in two consecutive stages. Then, a real-time model is developed by modifying the SBTR model in four stages. Finally, the braking control system is introduced in the model. The implementation of such advanced methodology will help to develop and verify the proposed solutions at initial design phases by reducing probable design faults.

## 3 Scaled bogie test rig

A scaled bogie test rig as shown in Fig. 3 was developed in the Centre for Railway Engineering, Central Queensland University, Australia. The test rig is the 1:4 scaled version of a freight bogie/wagon of 26.5 t axle load for 1067 mm narrow gauge track and LW3 wheel profile. The scaling factor ($\partial$) followed in this project is Iwnicki similitude method which is explained in Ref. [26]. The snapshot of the scale factor implemented for various parameters in this project is provided in Table 1. However, the scale parameter comparison between the actual bogie and SBTR is not the scope of this paper. The test rig consists of one bogie frame, two wheelsets, and four rollers.

### 3.1 Scaled bogie test rig model

A scaled bogie test rig is modeled as shown in Fig. 4 to replicate the physical scaled test rig. The parameters for the scaled bogie test rig (SBTR) model were derived from the actual test rig model developed in Gensys [27]. Additionally, the moment of inertia of the scaled model was derived from the SolidWorks model of the SBTR. All the components are modeled as rigid mass bodies with six degrees of freedom with some constraints as listed in Table 2. The scaled test rig is shown in Fig. 3, and basic parameters are presented in Table 3.

The connection between masses is shown in Fig. 4 with the number of DOFs indicated. The connection between the car body and bolster includes 10 couplings made up of

- two vertical coil springs with longitudinal, lateral, and vertical stiffness;
- one anti-roll stiffness element, two lateral bumpstops, one lateral damper, two vertical viscous dampers;
- one damping element and one stiffness element working in parallel in the longitudinal direction for traction rod.

The connection between bolster and side frames includes 14 couplings made up of

- six dampers and six stiffness elements for longitudinal, lateral, and vertical directions;

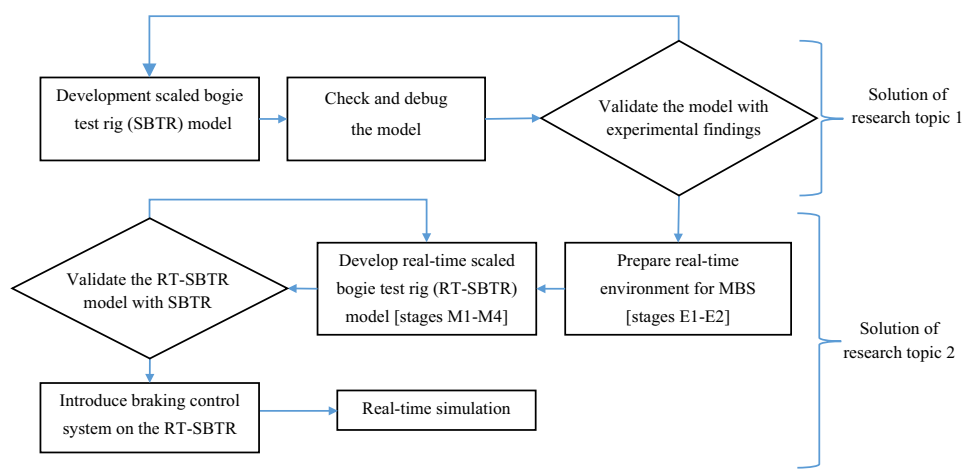

**Fig. 2** Methodology to solve the research questions

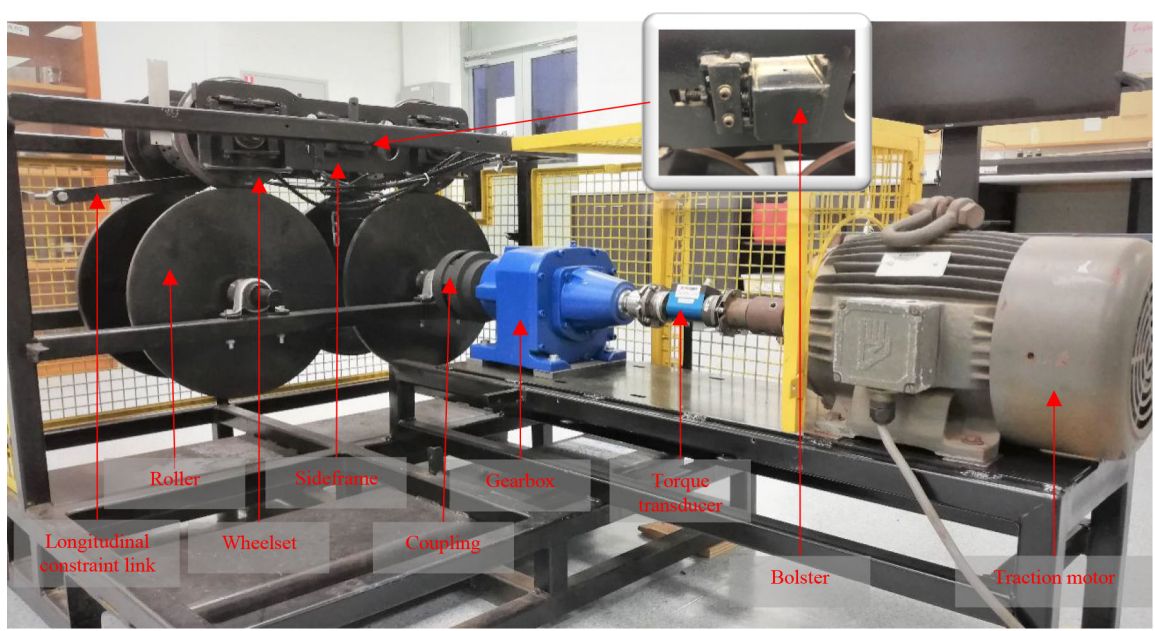

**Fig. 3** Scaled bogie test rig in Centre for Railway Engineering, Centre Queensland University Australia

- two yaw dampers.

Each wheelset is modeled as a single mass with six DOFs. The connection between two side frames and two wheelsets includes 24 couplings made up of

- twenty-four stiffness and damping elements for longitudinal, lateral, and vertical directions;
- four vertical and four lateral bumpstops.

Due to the finite vertical curvature of the roller as compared to real track (i.e., rail), there exist differences between wheel–rail and wheel–roller contact. In [28], the authors report how those two types of contact differ with regard to geometry, creep coefficient, stability, vibration response, and curve simulation. The difference in contact patch formation and the distribution of normal and tangential stresses are studied in [29]. In general, the vertical curvature needs to be considered in wheel–roller contact, unlike wheel–rail contact where such curvature is huge. The finite curvature of the roller makes the contact surface semi-axis shorter in the longitudinal direction which results in the small size of the contact patch area. The standard contact coupling *wr_coupl_polach* has been used in Gensys addressing the issues mentioned above [30].

**Table 1** Scaling factor considered

| Parameter | Scale factor |
| --- | --- |
| Scale | $1/\partial$ (1/4) |
| Dimension | $1/\partial$ (1/4) |
| Area | $1/\partial^2$ (1/16) |
| Volume | $1/\partial^3$ (1/64) |
| Mass | $1/\partial^3$ (1/64) |
| Velocity | $1/\partial$ (1/4) |
| Acceleration | $1/\partial$ (1/4) |
| Force | $1/\partial^4$ (1/256) |
| Torque | $1/\partial^5$ (1/1024) |
| Stiffness | $1/\partial^3$ (1/64) |
| Damping | $1/\partial^3$ (1/64) |
| Moment of inertia | $1/\partial^5$ (1/1024) |

### 3.2 Model-checking and debugging

The coefficient of friction in the wheel–rail contact patches is assumed to be 0.42 considering dry weather conditions [31]. The procedure followed in this stage is based on the Gensys online documentation [32, 33] as well as procedures provided in Table 1 of [34].

The bogie model was checked for syntax errors by using Gensys program RUNF_INFO [32] to analyze model code. A visual check was accomplished using the GPLOT [33] utility by plotting the bogie model in three dimensions and checking for faults such as inappropriate connection mounts and other geometrical errors. The quasistatic analysis was performed to analyze suspension response due to displacements of 1 cm in lateral (right) and vertical (downward) directions. In vertical displacements, the wheel loads increased evenly throughout all axles. Similarly, with 1 cm lateral displacement, the bogie was yawed with respect to the roller.

A time-stepping analysis was conducted to determine the critical speed at a normal time step of 1 ms. The initial speed of 67 km/h was implemented with deceleration at 3 km/h per second. Initial excitation was applied to the car body to introduce hunting. The bogie stops hunting at a speed of approximately 35 km/h as shown in Fig. 5 which corresponds to 140 km/h full-scale speed.

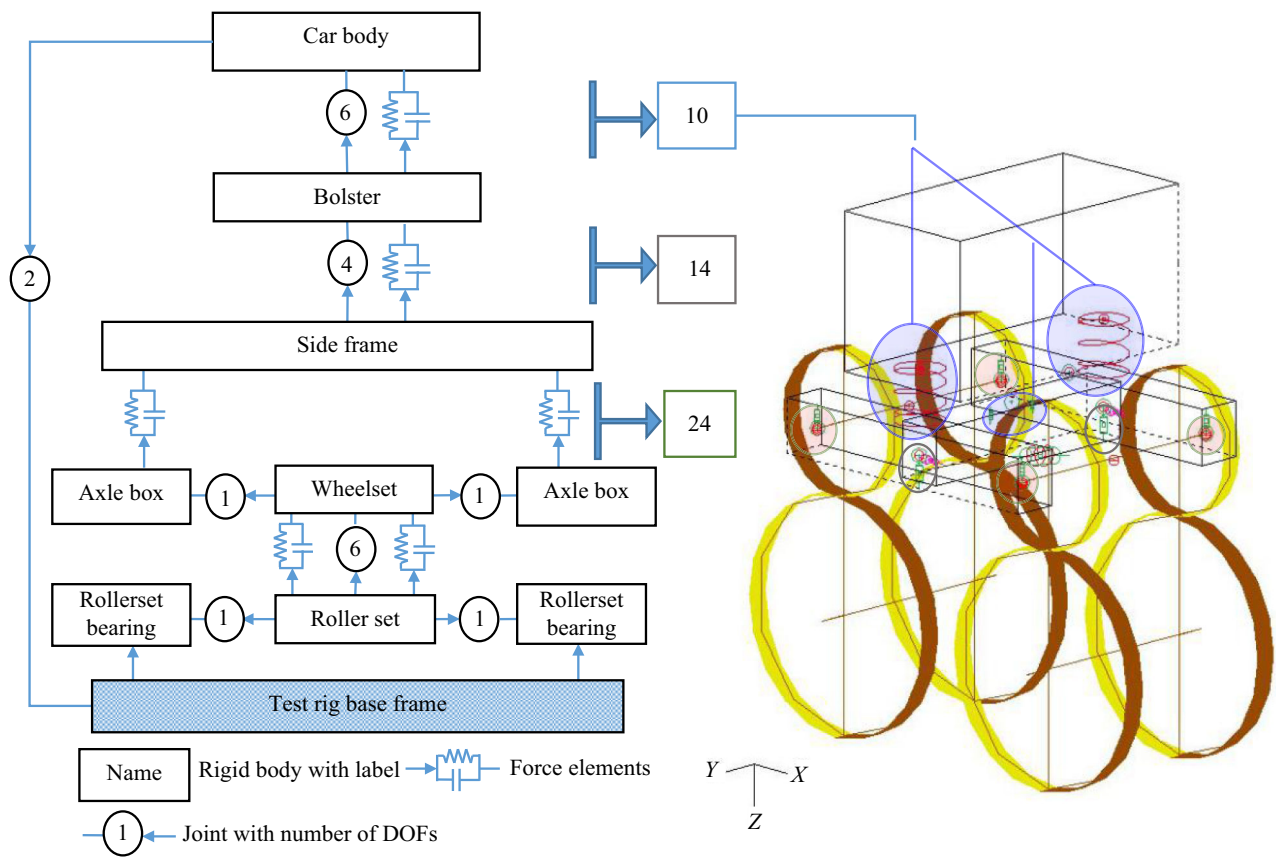

**Fig. 4** Bogie test rig model with mass, force element, constraints, and joint with the degree of freedom. Adopted from [27]

**Table 2** Constraints on bodies

| Degree of freedom | Longitudinal (x) | Lateral (y) | Vertical (z) | Roll (f) | Pitch (k) | Yaw (p) |
|---|---|---|---|---|---|---|
| Car body | Yes, $x=0$ | Yes | Yes | Yes, $f=0$ | Yes, $k=0$ | Yes |
| Bolster frame | Yes | Yes | Yes | Yes | Yes | Yes |
| Side frame | Yes | Yes | Yes | Yes | Yes | Yes |
| Axle | Yes | Yes | Yes | Yes | Yes, $k=0$ | Yes |
| Roller | No | No | No | No | Yes | No |

$x=0$, $f=0$, and $k=0$ refer to longitudinal translation displacement, roll, and pitch rotations being fixed to be equal to zero

**Table 3** Bogie mass and inertia parameters

| Component | Parameter | | | | |
|---|---|---|---|---|---|
| | Mass (kg) | Center of gravity, vertical (m) | Moment of inertia (kg·m$^2$) | | |
| | | | Roll | Pitch | Yaw |
| Car body | 86.8 | 0.20 | 450 | 200 | 450 |
| Bolster frame | 4.45 | 0.11 | 0.14 | 0.015 | 0.14 |
| Side frame | 5.10 | 0.115 | 0.15 | 0.02 | 0.17 |
| Axle | 25.10 | 0.115 | 0.65 | 0.28 | 0.65 |
| Roller | 40.21 | 0.20 | 1.66 | 1.17 | 1.66 |

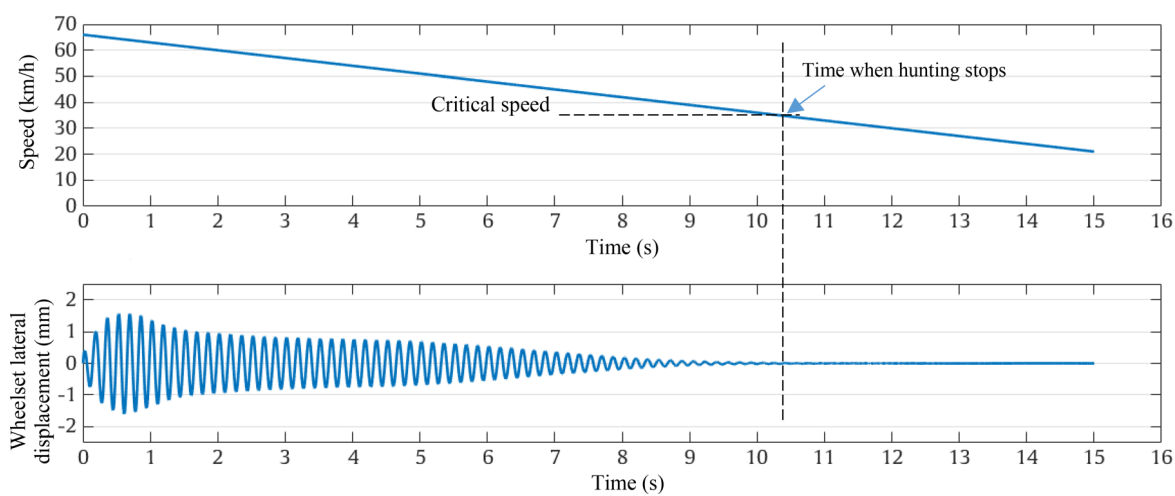

**Fig. 5** Hunting from lateral displacement in critical speed

### 3.3 Validate with the previous finding

The model is compared with the experimental results from the physical scaled bogie test rig. The test rig was tested for six different test scenarios (see Fig. 6). First, it was tested at a speed of 35 km/h with and without initial lateral displacement. Second, it was tested at 54.6 km/h with and without initial lateral displacement. Finally, it was tested at 67 km/h without initial lateral displacement. The initial lateral displacement was developed on the front wheelset by applying an external force for 100 ms, with a 427 N load being applied in all cases to match the experimental conditions. In the first case, the bogie stabilized in all conditions. For the second case, the bogie showed a hunting motion when initial lateral displacement was applied. For the third case, the bogie showed self-hunting motion after running for 5.5 s.

The second set of tests was accomplished to determine the maximum lateral displacement of the wheelsets when 850 N lateral force was applied on the wheelset. Three different cases were compared as shown in Fig. 7. Forces were applied on both wheelsets in the first case. In the second and third cases, the force was applied on the rear wheelset and front wheelset, respectively. The maximum

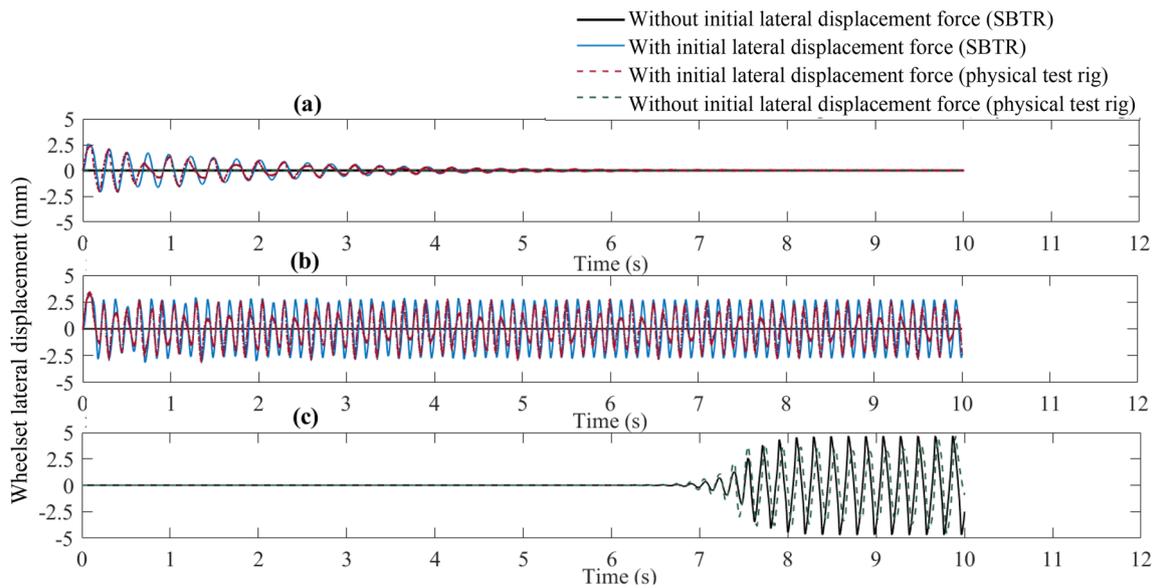

**Fig. 6** Lateral displacement at speeds of: **a** 35 km/h, **b** 54.6 km/h, and **c** 67 km/h

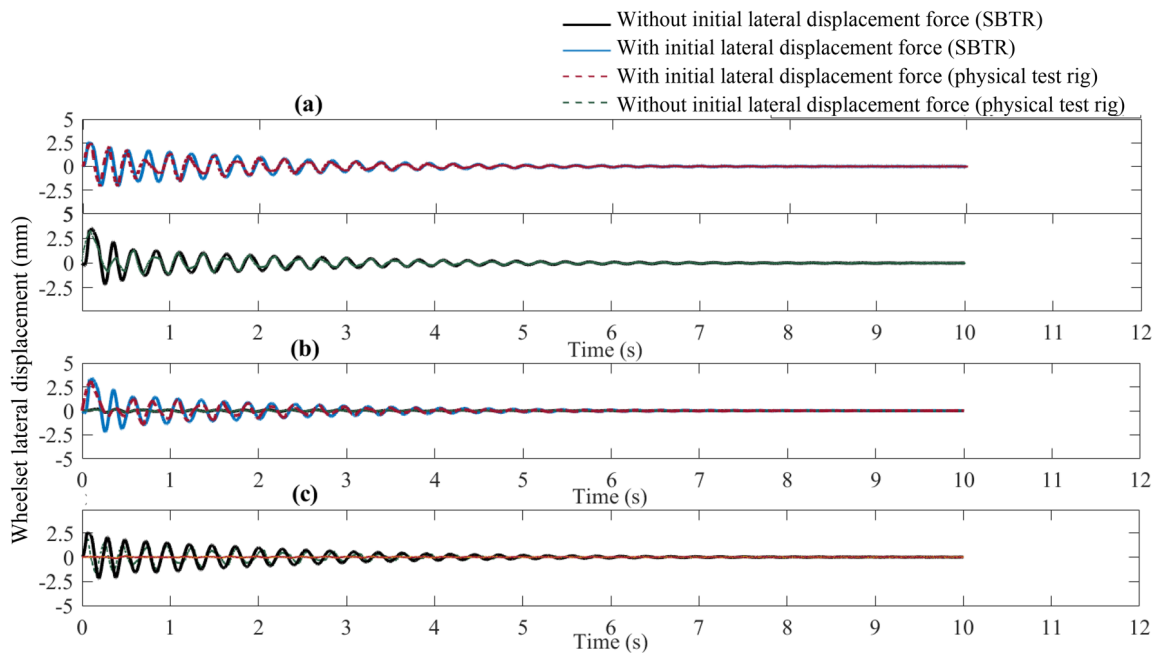

**Fig. 7** Lateral displacement on wheelset at 35 km/h with 850 N external lateral force applied on **a** both wheelsets, **b** rear wheelset, and **c** front wheelset

lateral displacement of the front wheelset was found to be 3.50 mm which was 0.25 mm more than the corresponding experimental result. Furthermore, the maximum lateral displacement of the rear wheelset was found to be 2.5 which is equal to the corresponding experimental result.

## 4 Real-time simulation of the model

The SBTR model is unable to replicate the physical test rig in the time domain. To be able to imitate the characteristics of its physical counterpart, the RT-SBTR is modeled by modifying the SBTR in four phases. For real-time simulation, a dedicated simulator environment is also created.

## 4.1 Calculation time in MBS

The test rig model consists of one bogie frame, two wheelsets, and four rollers. All the components are modeled as rigid mass bodies connected by force elements (i.e., force couplings) and constraints. In every simulation step, there are many other parameters and each of them has a unique task associated with it. Those parameters are: lsys defines the local coordinate system regarding the global coordinate system, coupl determines the coupling forces, func is the function in model script, mass creates mass inertia in the model, cnstr evaluates constraints, integ is the numerical integrator, and ds represents the data storage procedure. A special time estimator for each parameter has been implemented in Gensys which are represented by the parameter subscript of $t$. The total time $t_\text{tout}$ in Gensys for each output step is [25, 35]:

$$t_\text{tout} = t_\text{lsys} + t_\text{coupl} + t_\text{func} + t_\text{mass} + t_\text{cnstr} + t_\text{integ} + t_\text{ds}, \quad (1)$$

where $t_\text{lsys}$ is the computational time spent on the position definition of local coordinate systems with reference to the global coordinate system, $t_\text{coupl}$ the computational time spent on commands for coupling elements (coupling elements are elements of various types that connect masses to each other), $t_\text{func}$ the computational time required for the calculation of defined functions in the model script, $t_\text{mass}$ the computational time spent on mass commands (a mass command creates inertia in the model, e.g., car body, bogie, wheelset, etc.), $t_\text{cnstr}$ the computational time spent on constraint commands, $t_\text{integ}$ the computational time required for calculation inside of the numerical integrator, and $t_\text{ds}$ the computational time required for output data storage.

## 4.2 Preparation of real-time environment for MBS simulation

For a valid real-time simulation, the real-time simulator must accurately execute the output within the same time duration that its physical equivalent would. A dedicated simulator environment is required to perform such simulation. Two different environment preparation stages, namely E1 and E2, as shown in Fig. 8 have been collectively implemented to improve the calculation time and satisfy the real-time simulation requirement [36].

In stage E1, the real-time kernel is implemented over a generic kernel. The Gensys MBS runs under the UNIX environment. The real-time operating system (RTOS) kernel uses a preemptive-based scheduling algorithm which allows the scheduler to forcibly perform a context switch to execute the desired high priority process without waiting for kernel function to complete its execution. In context switch, the state of a process or a thread is stored which can be restored and resumed from the same stage later. It allows sharing the single CPU among multiple processes. The symmetric multiprocessing (SMP) architecture of the kernel enables access to a single, shared main memory where two or more identical processors are connected. Since a CPU with multicore processors is used in this project, the SMP architecture applies to the cores, considering them as separate processors.

In stage E2, the RAMdisk software is used. The time taken for data storage $t_\text{ds}$ can be reduced by using RAMdisk. The computer's RAM is still faster than even modern solid-state drives. RAM can be used as a lightning-fast virtual drive known as RAMdisk. This program would reserve a section of RAM. All the files on the disk would be stored in your RAM (see Fig. 9c). It could help to optimize performance because the load times of the installed programs in a RAMdisk have near-zero latency because those data would already be stored in the fastest memory possible. Save a file also happens almost instantly as it would just be copied to another portion of RAM. This would mean RAMdisk speeds up the application load times and file read/write times for files saved in the RAMdisk. To automatically create the RAMdisk at every boot, the automounting option was enabled. However, RAM is volatile memory and the content of the RAM can be lost if the computer loses power. To deal with such nonpersistent memory, a regular backup was set up by creating a bash script to allow the periodic backup every 10 min in this work.

Another option that might be useful is RAM-based solid-state drives (RAM-SSD). These are solid-state drives that contain RAM instead of typical flash memory. They are much faster to read and write as compared to RAM. Such drives contain a battery so that they can maintain the contents of the RAM even if the computer loses power. Since RAM is more expensive than flash memory, RAM-SSD is an expensive option.

## 4.3 Preparation of real-time scaled bogie test rig (RT-SBTR) model

The SBTR model was further simplified in the following stages, namely M1, M2, M3, and M4, as presented in

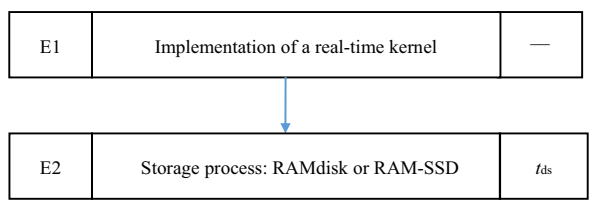

**Fig. 8** Stages to prepare a real-time environment for MBS simulation. In each stage, the columns from first to third represent name, task, and associated time of that stage, respectively

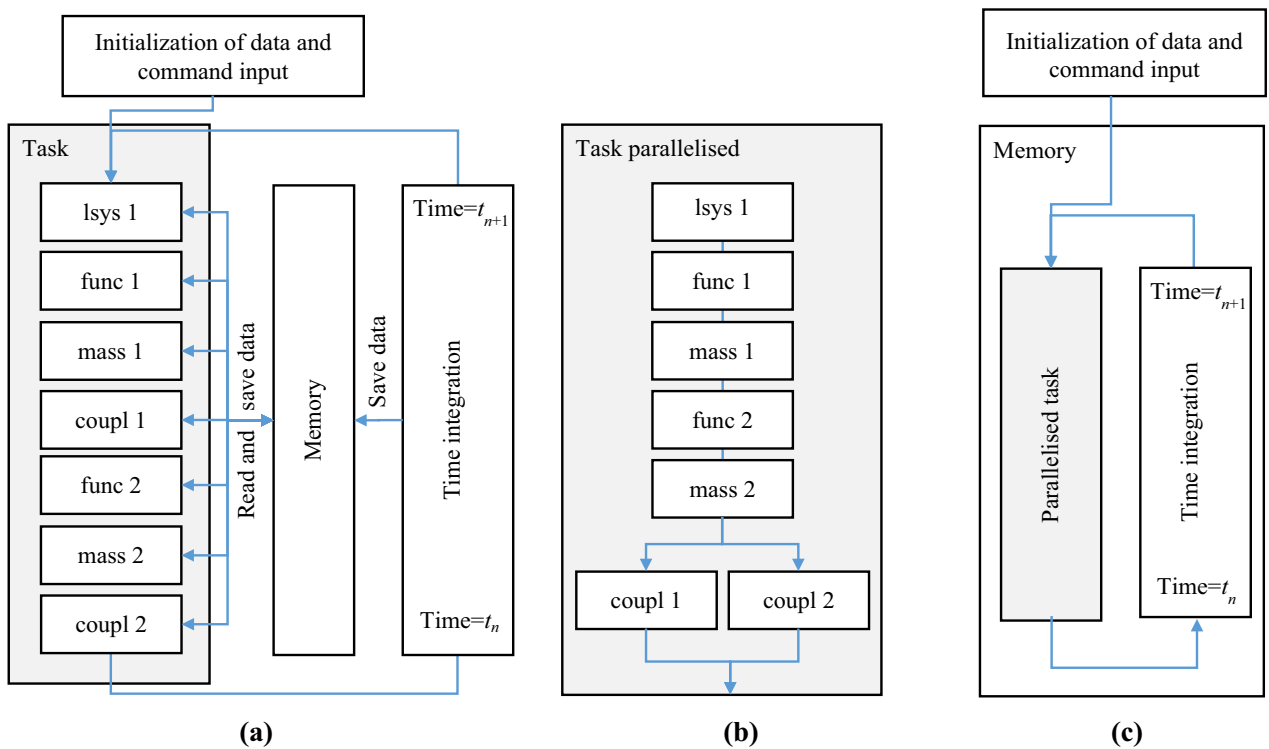

**Fig. 9** Time-domain simulation in Gensys: **a** original serial computing-based numerical steps, **b** parallel computing-based numerical steps, **c** parallel computing using RAMdisk

**Fig. 10** Stages to model RT-SBTR. In each stage, the columns from first to third represent name, task, and associated time of that stage, respectively

Fig. 10 to obtain a model suitable for real-time application with reasonable trade-offs with the accuracy. The execution time that is associated with tasks has been presented in the third column of Fig. 10. The significant reduction in the associated time leads to reducing the total time required (i.e., $t_{out}$) to execute in each step.

In stage M1, an appropriate numerical solver is chosen. The list of potential numerical solvers for real-time simulation in the Gensys MBS package is presented in [37]. The real-time solvers in Gensys and its MATLAB equivalent are presented in [25] for broader understanding. The SBTR has large numbers of friction elements and small masses. For accurate numerical simulation of such a model, it requires a more dedicated numerical solver such as Runge–Kutta which is a four-step method with a fixed step-size controller. For a simplified model in this study, the two-step Runge–Kutta (Heun) numerical integrator has been chosen which ignores the backstepping.

In stage M2, a simplified contact model is considered. The full model is capable of 3-point contacts. For a simplified model, a single-point contact is considered. Another reason for considering a single point is that 1 point enables only one wheel–roller force coupling which can be assigned to 1 CPU core. Thus, wheel–roller force coupling of a bogie can be assigned via 4 CPU cores simultaneously to allow parallel computing. The contact subroutine can be further simplified by using a lookup table which is not considered in this project [38].

In stage M3, multiple stiffnesses and damping elements in couplings are replaced by equivalent stiffness and damping elements. With the reduction in elements, the number of couplings and functions were reduced. Similarly, graphical figures, saving variables, and the associated post-processing section were removed after the initial check. Thus, the functions and storage were reduced. As a result, the total calculation time was remarkably decreased.

In stage M4, parallel computing is implemented. Parallel computing is not attempted to simulate the bogie model in railway vehicle engineering. This attempt ensures faster execution of the program. Parallel computing was enabled by using the OpenMP [39]. In serial computing (see Fig. 9a), the parameters are calculated in a series arrangement. In parallel computing (see Fig. 9b), the normal components are executed in series and all the remaining parallelized components are executed simultaneously at the end of every time step. In this project, the wheel–roller coupling forces are parallelized. Thus, parallel computing reduces the total time taken by reducing the time taken to execute the coupling force (i.e., $t_{\text{coupl}}$). Due to the availability of only four CPU cores, parallel computing was implemented on the wheel–roller force couplings which are the most execution-time demanding force couplings (see Fig. 11). Upon available of more CPU cores, other parallelizable force couplings can also be parallelized. Activating parallel computing in Gensys has to be compiled with fopenmp flag.

### 4.4 Brake control system

The wagon model is then incorporated with a control system. The brake controller uses a feedback control approach. The detail of the brake controller is explained in [1]. The schematic of the braking control system implemented in the model is presented in Fig. 12, where $\lambda$ denotes the calculated slide, $\lambda_o$ the reference slide, $T_b$ the brake torque, $T_a$ the adhesion torque, $r$ the nominal rolling radius of the wheel, $F_a$ the adhesion force, $v$ the linear roller speed, and $\omega$ the angular wheelset velocity.

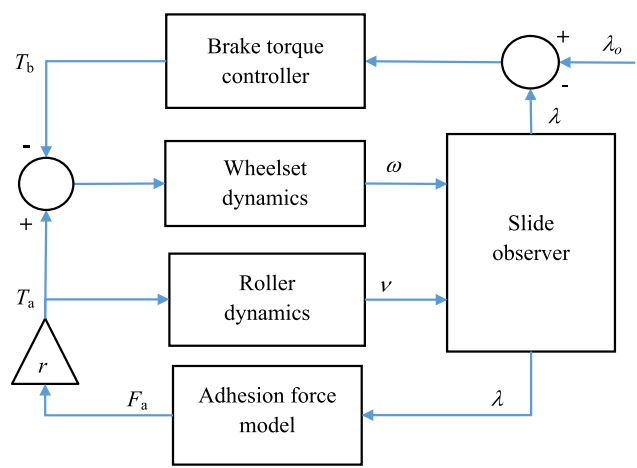

Fig. 12 Schematic diagram of the wagon model with brake controller

The brake controller is a proportional–integral controller which uses the slide error as the input signal. A first-order lag filter is used to compensate for the associated delay of the brake system. The slide is estimated from the following relationship:

$$\lambda = 1 - \frac{\omega r}{v}. \qquad (2)$$

## 5 Simulation and results

In this section, the RT-SBTR is validated with SBTR. Then, the computational improvement in each stage during the real-time environment preparation and real-time model preparation is discussed.

### 5.1 Validation of the real-time SBTR model

The verification of the RT-SBTR with respect to SBTR has been done by the critical speed test and vertical contact force test. For the critical speed test, the initial speed was set to 67 km/h and reduced at the rate of 3 km/h per second. The critical speed of the RT-SBTR was found to be approximately 31 km/h as compared to nearly 33 km/h for SBTR as can be seen from Fig. 13. The reason for the slightly larger critical speed of the RT-SBTR is due to the simplification of the model. The vertical contact force test of the RT-SBTR was conducted to test the stability of the numerical solver. Forces from both SBTR and RT-SBTR showed similar characteristics and stabilized in a short time.

### 5.2 Case studies

In this section, three case studies are simulated and compared. In Case 1, simulations based on the different stages followed to prepare a real-time environment as mentioned

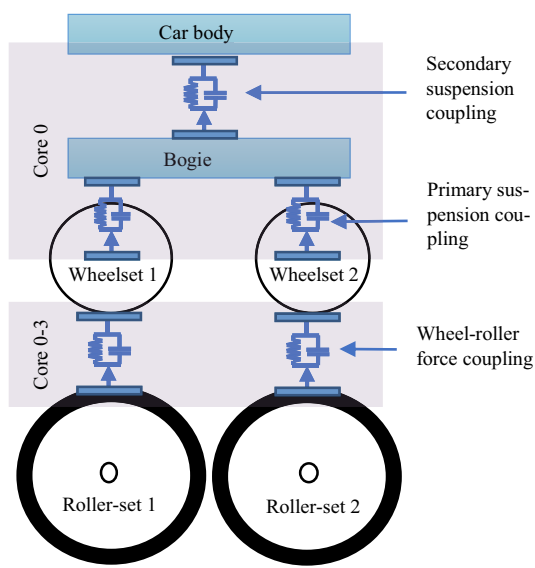

Fig. 11 Assignment of CPU core on the wheel–roller force coupling elements

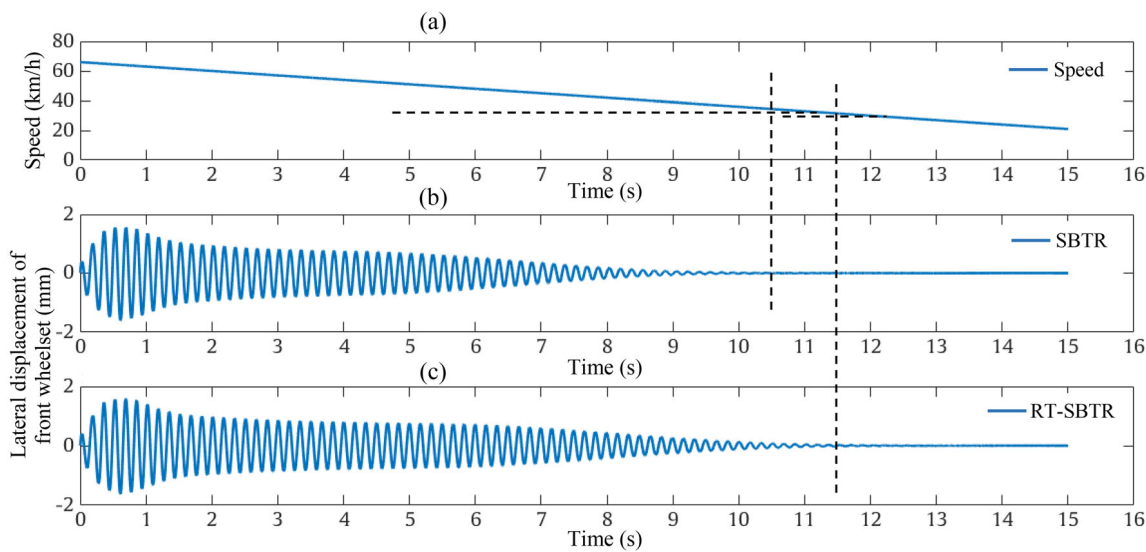

**Fig. 13** a Critical speed, lateral displacement of wheelsets at critical speed for **b** SBTR, and **c** RT-SBTR

in Sect. 4.2 are performed. In Case 2, simulations based on the different stages followed to model the RT-SBTR as explained in Sect. 4.3 are shown. In Case 3, the simulation of RT-SBTR model incorporating the brake control system is presented.

#### 5.2.1 Case 1

For all simulations, the SBTR model is simulated for 1 s. The integration time step was set fixed at 0.25 ms, and the results were saved every 1 ms. The first simulation was conducted with Gensys installed on the generic Linux kernel. Then, the simulations were conducted at every stage as explained in Sect. 4.2. Figure 14 presents the total time taken to execute the simulation of 1 s.

The computing time also indicates how many times the simulations are slower than real time. For example, the original computing time is 3.99 times slower than real time. In the first stage of improvement, the computing time is improved by 2.55 times with the implementation of the real-time Linux environment. Since the simulated model is not memory intensive, not many differences were seen in the second stage.

However, a clear difference in performance can be seen if a complete railway vehicle is simulated. The total time $t_{tout}$ in Gensys is presented in Fig. 15. In the RT operating environment, the output has less jitter as compared to the non-RT environment. Overall, the computing time was improved by 2.57 times.

#### 5.2.2 Case 2

This simulation is based on the different stages followed to develop RT-SBTR from SBTR mentioned in Sect. 4.3. For all simulations, the RT-SBTR model is simulated for 1 s. The integration time step was set fixed at 0.25 ms, and the results were saved every 1 ms. The simulation speed was improved by more than 600%, while preparing the real-time environment, however, it was not able to achieve the real-time speed.

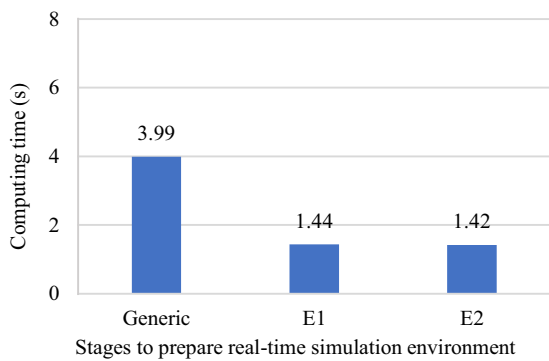

**Fig. 14** Total computing time to execute the simulation of 1 s

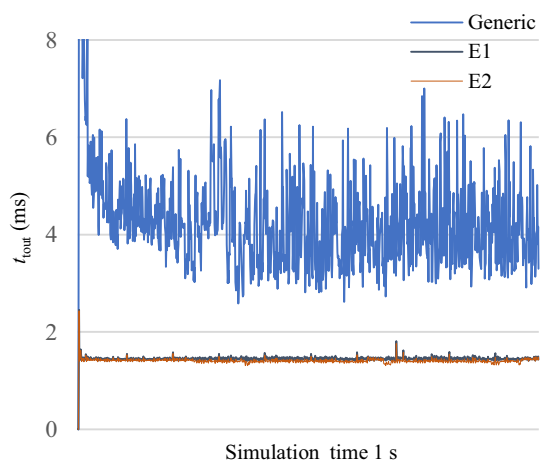

**Fig. 15** Total time $t_{tout}$ from the simulation in every stage with a timestep of 1 ms for case 1

To achieve real-time speed, the simulation needs to be executed faster than 1 s in this work. Thus, in the second stage, the scaled bogie test rig (SBTR) model is improved in four stages to develop the RT-SBTR model. Figure 16 shows the total time taken to execute the simulation of 1 s. The total time $t_{tout}$ in Gensys is presented in Fig. 17. Firstly, a two-step Runge–Kutta (Heun) numerical integrator has been chosen which ignores backstepping. It improved the computing speed of 1.42 s by 42% to reach 0.82 s. Secondly, single-point wheel–roller contact was considered in the RT-SBTR model which improved the speed by 17%. In the third stage, multiple stiffnesses and damping elements in couplings were replaced by single equivalent stiffness and damping elements. Similarly, graphical figures, saving variables, and associated post-processing section were also removed after the initial check. This improved the computational speed by 29% to reach 0.48 s. Finally, parallel computing has been applied to the wheel–roller force couplings which are the most execution-time demanding force couplings. Since single-point wheel–roller contact is considered in this work, the total 4 wheel–roller force couplings were assigned to the 4 CPU cores available to allow parallel computing. Thus, the computational speed of the final RT-SBTR model was found to be 2.5 times faster than the real time.

### 5.2.3 Case 3

A braking control system is also incorporated in the model. Two different slip references have been chosen: 0.1 for dry and 0.2 for wet conditions. The wet friction condition has been chosen for the first 4 s, and the friction condition remains dry for the next 4 s. The brake is applied when the speed is 5.56 km/h.

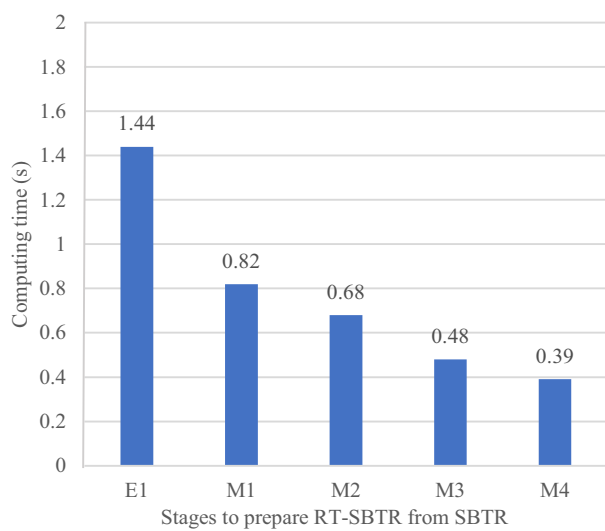

**Fig. 16** Total computing time to execute the simulation of 1 s at different stages of preparing RT-SBTR from SBTR

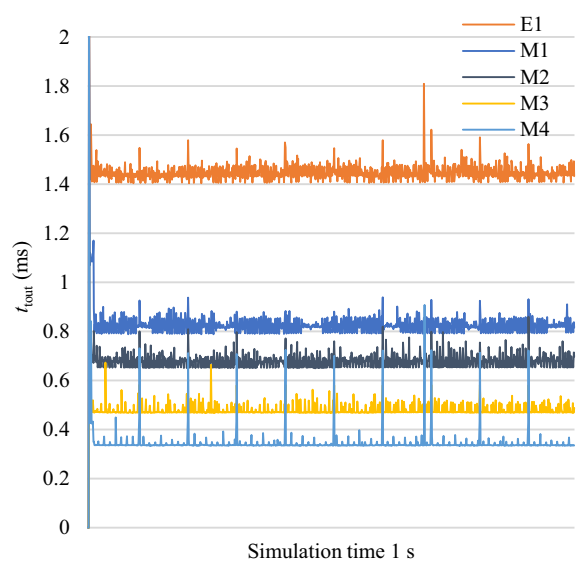

**Fig. 17** Total time $t_{tout}$ from the simulation in every stage with a timestep of 1 ms for case 2

The results obtained from the real-time simulation for wheelset longitudinal slip, adhesion coefficient, speed, and calculation time for wet and dry friction conditions can be seen in Fig. 18. Additionally, the wheelset slip is slightly higher than the reference slip provided. The adhesion coefficient increased from 0.11 to 0.19 when the friction condition shifts from wet to dry. Even with the brake control system implemented, the computational time is faster than real time.

## 6 Discussion and conclusion

Most of the railway wheel–rail adhesion studies have been conducted using simplified designs which may not be accurate. As an alternative, the complex system of a full vehicle model provides accurate results but may create difficulty in preparing its design. Thus, a bogie model has been selected in this project which has a slight trade-off with accuracy to significantly reduce complexity in design. Furthermore, to exactly represent the physical bogie test rig in the time domain, the simulation model needs to run faster than the physical test rig. In other words, the simulation model needs to be capable of performing in a discrete time with fixed step by solving the internal state equations and functions representing the physical counterpart system in less than actual time. In general, such a model is known as a real-time model. An important question then arising is how

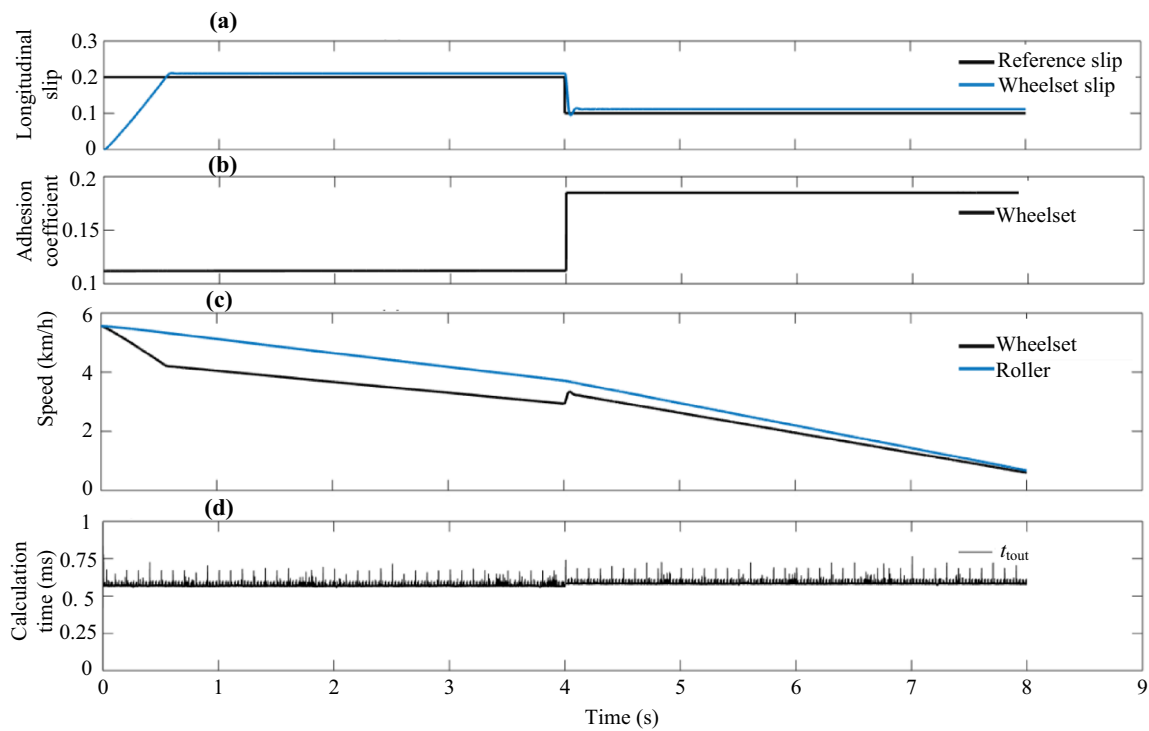

**Fig. 18** Parameters in time domain: **a** reference slip and an estimated longitudinal slip of wheelset, **b** adhesion coefficient, **c** speed of the wheelset versus roller, and **d** total time for each output step of 1 ms

to achieve such a model which has higher computing speed, but also good modeling accuracy. To address this question, the model has been prudently simplified, and then, an innovative computing scheme involving parallel computing has been implemented.

Three simulation cases were carried out for 1 s each for comparison purposes. In case 1, the computational speed while preparing a real-time simulation environment was recorded. The computing speed was improved by 2.57 times by establishing the simulation environment without changing the model. However, the time taken to simulate was far from real time. In case 2, the model was further simplified, and parallel computing was implemented. This reduced the computation time from 1.42 s to 0.39 s which was 2.5 times faster than real time. In both cases, the control system was not incorporated into the model. In general, the addition of the control system makes the model more computably expensive and unstable. In this work, the computational time of the model with the control system included is 0.62 s which is still faster than real time. Thus, the proposed technique can satisfy the requirements of the real-time simulation of the system. However, the brake control used in this work is simplified and a more precise control scheme may lead to a higher computational commitment. A computer equipped with Intel® Core™ i5-4570 CPU @3.20 GHz with 8 GB of RAM has been used for the simulation process with the Gensys MBS software. There are more dedicated and powerful computers already available in the market, which may help to achieve faster computational speed.

Simulation modeling of a scaled test rig for an adhesion study at the wheel–rail interface is described in this paper. To progressively conduct the study of the complexity of the modern railway vehicle, hardware-in-loop simulation integrating the physical test rig will be performed in future.

**Acknowledgements** The authors greatly appreciate the financial support from the Rail Manufacturing Cooperative Research Centre (funded jointly by participating rail organizations and the Australian Federal Government's Business Cooperative Research Centres Program) through Project R1.7.1—"Estimation of adhesion conditions between wheels and rails for the development of advanced braking control systems." Tim McSweeney, Adjunct Research Fellow, Centre for Railway Engineering is thankfully acknowledged for his assistance with proofreading.